\documentclass[notimes]{georeport}

\usepackage{natbib}
\usepackage{graphicx}
\usepackage{color}
\usepackage{listings}
\usepackage{amsmath}
\usepackage{amssymb}
\usepackage{amsbsy}
\usepackage{float}
\usepackage{wasysym}
\usepackage{setspace}
\usepackage{bm}

      \newcommand{\fl}{f_{\lambda}}
      \newcommand{\dl}{d_{\lambda}}
      \newcommand{\bx}{{\bf x}}

\setfigdir{Fig}

\begin{document}
\title{Full Waveform Inversion by Source Extension: Why it works}
\author{William. W. Symes \thanks{The Rice Inversion Project,
Department of Computational and Applied Mathematics, Rice University,
Houston TX 77251-1892 USA, email {\tt symes@caam.rice.edu}.}}

\lefthead{Symes}

\righthead{Why it works}

\maketitle
\begin{abstract}
  An extremely simple single-trace transmission example shows how an
  extended source formulation of full waveform inversion can produce
  an optimization problem without spurious local minima (``cycle
  skipping''). The data consist of a single trace recorded at a given
  distance from a point source. The velocity or slowness is presumed
  homogeneous, and the target source wavelet is presumed
  quasi-impulsive or focused at zero time lag. The source is extended
  by permitting energy to spread in time, and the spread is controlled
  by adding a weighted mean square of the extended source wavelet to
  the data misfit, to produce the extended inversion objective. The
  objective function and its gradient can be computed explicitly, and
  it is easily seen that all local minimizers must be within a
  wavelength of the correct slowness. The derivation shows several
  important features of all similar extended source algorithms. For
  example, nested optimization, with the source estimation in the
  inner optimization (variable projection method), is essential. The
  choice of the weight operator, controlling the extended source
  degrees of freedom, is critical: the choice presented here is a
  differential operator, and that property is crucial for production
  of an objective immune from cycle-skipping.
\end{abstract}

\section{Introduction}
Full Waveform Inversion (FWI), or estimation of earth structure by
model-driven least squares data fitting, is now well-established as a
useful tool for probing the earth's subsurface
\cite[]{VirieuxOperto:09,Fichtner:10}. However, so-called ``cycle-skipping'', the tendency of iterative FWI
algorithms to stagnate at suboptimal and geologically uninformative
earth models, still impedes its use. Because the computational size of field inversion tasks
is very large, only iterative local (descent) minimization of the data
misfit function is computationally feasible. However local
descent methods avoid suboptimal stagnation only if initial models are
already quite close to optimal, in the sense of predicting the arrival
times of seismic events to within a small multiple of a dominant
wavelength \cite[]{GauTarVir:86,Plessix:10}.

This paper concerns one of the several ideas that have been advanced to
overcome cycle-skipping, namely so-called extended inversion
\cite[]{geoprosp:2008}. ``Extended'' signifies that addional degrees
of freedom are provided to the modeling process, in the hope of
opening up more effective routes to geologically informative models
with acceptable data fit. Since these extended degrees of freedom are
not part of the basic physics chosen to model the data acquisition
process, they should be suppressed in the eventual solution. Extended
inversion methods differ by the choice of additional degrees of
freedom, and by choice of penalty applied to eliminate them in the final
result.

Many of these extended inversion concepts sound plausible, and appear
to work at least to some extent as one might hope from their heuristic
justifications. However very few of these approaches have been
underwritten by mathematical argument: in essence, they are mostly
justified only ``in the rear-view mirror'', with no assurance that
failure is not just around the corner, at the next example. On top of
that, some of these approaches, for example those based on the
computationally attractive Variable Projection Method (``VPM'') of
\cite{GolubPereyra:03}, are cast in such form that the reasons for
success are not readily apparent.

This note shows exactly how VPM leads to successful velocity updates
for a particular extended inversion approach to a very simple inverse
problem, which asks that a homogeneous velocity field be deduced from
one trace at known offset. I put forward this inverse problem and
extension-based solution not because there are not simpler
ways of answering the question it poses - there certainly are - but
because the formal ingredients of waveform-based velocity estimation
in this very simple setting are common to many similar extended
inversion algorithms, and because in this case every computation can
be done analytically, nearly to completion. In particular, it becomes
clear why the VPM gradient formula produces a constructive update,
with no possibility of stagnation away from the global minimum.

The extended inversion approach
developed here uses a {\em source extension}, in which source
parameters form the additional degrees of freedom. This type of
extension presumes that the actual or target source is constrained in
some way; the extended source is allowed to violate the
constraint. For recent overview of source extension methods, see
\cite{HuangNammourSymesDollizal:SEG19}. Source
extension methods have computational complexity approximately the same
as that of FWI, a signal advantage over the alterantive medium
extension class described for instance in \cite[]{geoprosp:2008}.

For the problem considered here, the source model amounts 
to a wavelet, and the target wavelet is assumed to be non-zero only in
a short time
interval (an approximate impulse, perhaps as the result of signature deconvolution). The extension consists in permitting energy to spread
in time at intermediate iterations of the inversion. A simple penalty
for energy spread (second moment of square amplitude) drives the
extended source towards a focused source approximately satisfying the assumed
constraint. Not all penalties are created equal: the penalty used here
has the critical {\em (pseudo-)differential} attribute necessary for
avoidance of cycle-skipping, as will be explained in the Discussion section.

%despite not incorporating explicitly the extension measure
%(``annihilator'') central to the approach.

I begin with a quick sketch of constant density acoustics,
and describe the single-trace transmission inverse problem. To make
the role of data frequency content clear, I introduce a family of
noise-free data parametrized by wavelength. For completeness, I show how
the standard FWI approach to this problem generates multiple local
minima that will be found by any descent method unless the initial
estimate predicts travel time from source to receiver with an error
on the order of a wavelength. The next section describes the source
extension objective, and the reduced objective
produced by VPM. As VPM eliminates the extended source, this
function depends only on the velocity, just as does the FWI objective.  A
nearly-explicit calculation of the VPM gradient 
shows that the only stationary points are ``within a wavelength'' of
the correct velocity, used to build the data: that is, cycle-skipping
cannot occur. The paper ends with a discussion of the parallels
between the calculations presented here and the structure of other
extended inversion methods applicable to field-scale velocity
estimation, and the critical role that the differential nature of the
extension penalty plays in the success of this and other extension methods.

\section{Preliminaries}
Assume small amplitude (linearized) acoustic propagation, constant
density, and isotropic point source and receiver. Denote by $m(\bx)$
the slowness (reciprocal velocity) at spatial position $\bx$, $f(t)$
the time dependence of the point source (``wavelet'') at location
$\bx=\bx_s$. Then the (excess) pressure field $p(\bx,t)$ obeys a
scalar wave equation:
\begin{eqnarray}
  \label{eqn:awe}
  \left(m(\bx)^2\frac{\partial^2 p}{\partial t^2} - \nabla^2\right) p(\bx,t) &=&
                                                                         f(t)\delta(\bx-\bx_s) \nonumber\\
  p(\bx,t)&=&0, t\ll 0
\end{eqnarray}
Suppose that a single trace is recorded, at distance $r>0$ from the
source position $\bx_s$. The dominant information in a single
trace is the transient signal time of arrival, constraining only the mean slowness in the
region
between source and receiver, so assume that the
slowness is constant, that is, independent of position $\bx$. The pressure field is simply the  the source
wavelet $f(t)$ convolved with the
acoustic Green's function, for which an analytic expression is
available in the constant $m$ case \cite[]{CourHil:62}:
\begin{equation}
  \label{eqn:homsol}
  p(\bx,t) = \frac{1}{4\pi |\bx-\bx_s|}f\left(t-m|\bx-\bx_s|\right).
\end{equation}

The receiver location $\bx_r$ lies at distance $r$ from the source
location $\bx_s$, that is, $|\bx_r-\bx_s|=r$. The predicted signal at
$p(\bx_r,t)$ depends nonlinearly on the slowness $m$ and linearly on the
source wavelet $f$. Therefore it is naturally represented as the
action of a $m$-dependent linear operator $S[m]$ on $f$:
\begin{equation}
\label{eqn:mod}
S[m]f (t)= p(\bx_r,t) = \frac{1}{4\pi r}f\left(t-mr\right).
\end{equation}

Ignoring amplitude, this map implements a $m$-dependent time
shift. This time shift operator is the basis of many descriptions of
the cycle-skipping phenomenon (for example, \cite{VirieuxOperto:09},
Figure 7), so it is unsurprising that an analysis of cycle-skipping
can be based on the simple modeling operator described above, which
amounts essentially to a time shift. To make the link with wavelet
frequency content manifest, I introduce a family $\{\fl\}$ of
wavelets indexed by $\lambda$, a parameter having dimensions of time,
\begin{equation}
  \label{eqn:deffl}
  \fl(t) = \frac{1}{\sqrt{\lambda}}f_1\left(\frac{t}{\lambda}\right).
\end{equation}
The argument $s$ of the ``mother wavelet'' $f_1$ is
nondimensional. The only constraints placed on $f_1$ are that (i) 
$f_1(s)=0$ for $|s|\ge 1$, and (ii) $f_1$  has positive
mean-square, that is, does not vanish identically. Note that
the scaling is such that the mean-square
\[
  \|\fl\|^2 = \int\,dt\,|\fl(t)|^2
\]
is independent of $\lambda$.

I shall refer to $\lambda$ as ``wavelength'': if $f_1$ has a dominant
period of oscillation, then so does $\fl$, and it is proportional to
$\lambda$.

To this family of wavelets and a choice of target slowness $m_*$  corresponds a family of noise-free data
\begin{equation}
  \label{eqn:defdata}
  \dl=S[m_*]\fl.
\end{equation}
This family of data in turn defines a family of inverse problems, to
which I now turn.

\section{Full Waveform Inversion}
The preceding section provided all of the raw ingredients to define
full waveform inversion for estimation of $m$ from a single trace.
It is only $m$ that is to be determined: the $\lambda-$dependent family of wavelets
$\{\fl\}$ is regarded as known, along with the data family $\{\dl\}$. The aim
is to chose $m$ to minimize 
\begin{equation}
  \label{eqn:FWIfix}
  J_{\rm FWI}[m] = \frac{1}{2}\|S[m]\fl-\dl\|^2.
\end{equation}
for all values of $\lambda > 0$.

Written out in detail, this objective function is
\[
 J_{\rm FWI}[m] =  \frac{1}{32\pi^2
    r^2}\int\,dt\,\left|\fl\left(t-mr\right)-\fl\left(t-m_*r\right)\right|^2
\]
Since $f_1$ vanishes for $|t|>1$, $\fl$ vanishes for $|t|>\lambda$,
and $S[m]\fl$ vanishes if $|t-mr|>\lambda$. So if $|mr-m_*r|
= |m-m_*|r > 2\lambda$, then $|t-mr|+|t-m_*r| \ge |mr-m_*r| >
2\lambda$ so either $|t-mr|>\lambda$ or $|t-m_*r|>\lambda$, that is,
either $S[m]\fl(t)=0$ or $S[m_*]\fl(t)=0$. Therefore $S[m]\fl$ and
$S[m_*]\fl$ are orthogonal in the sense of the $L^2$ inner product:
\begin{equation}
  \label{eqn:ortho}
  |m- m_*|r > 2\lambda \,\, \Rightarrow \,\, \langle S[m]\fl,
  S[m_*]\fl\rangle = \int\,dt\,S[m]\fl(t)S[m_*]\fl(t) = 0
\end{equation}
But $\dl = S[m_*]\fl$, so this is the same as saying that $\dl$ is
orthogonal to $S[m]\fl$. So conclude that
\begin{equation}
  \label{eqn:iso}
  |m- m_*|r > 2\lambda \,\, \Rightarrow \,\, J_{\rm FWI}[m]=\frac{1}{16\pi^2
    r^2}\|f_1\|^2.
\end{equation}
using the previously observed independence of $\|\fl\|$ from
$\lambda$.

That is, for slowness $m$ in error by more than $2\lambda/r$ from the
target slowness $m_*$, the FWI objective $J_{\rm FWI}$ is perfectly
flat: all nearby values of $m$ are local minima. Therefore the local
exploration of the objective
gives no useful information whatever about constructive search
directions towards the global minimizer $m=m_*$, for which the
objective value is of course = 0, if the initial estimate of $m$ in
error by an amount greater than a fixed multiple of $\lambda$. This is
precisely the behaviour of FWI noted many times in the literature: the
model must be known to ``within a wavelength''. 

\section{Extended Source Inversion}
As mentioned in the introduction, the modeling operator introduced in
the last section, $m \mapsto S[m]\fl$, may be extended simply by
including the source wavelet as one of the model parameters: that is,
the model vector becomes $(m,f)$, and the modeling operator, $(m,f)
\mapsto S[m]f$.

The reader will have no trouble seeing that the data misfit using this
extended modeling operator can always be made to vanish entirely by
proper choice of wavelet $f$, unless $f$ must satisfy some additional
constraints. \cite{HuangNammourSymesDollizal:SEG19} describe a
plethora of possible constraints for this and similar source
extensions. Many (but not all) take the form of a quadratic penalty,
that is, the mean square of $Af$, $A$ being a suitable operator,
commonly dubbed an {\em annihilator}: in many examples the ideal
output for $A$ applied to a source obeying the target constraints
is the zero vector \cite[]{geoprosp:2008}. With an annihilator $A$, to
be chosen below, the penalty form of extended inversion is: minimize
over $\{m,f\}$
\begin{equation}
\label{eqn:obj}
J_{\rm ESI}[m,f] = \frac{1}{2}(\|S[m]f-\dl\|^2 + \alpha^2 \|Af\|^2).
\end{equation}

The choice of penalty weight $\alpha$ has a profound influence on the
character of this optimization problem. For the moment, I will mandate
only that $\alpha > 0$. A detailed discussion of methods for setting
$\alpha$ is beyond the scope of this brief paper.  In the Discussion
section, I mention an effective method for choosing $\alpha$ to
optimize convergence of iterative minimization, that applies (and has
been applied) to much larger-scale inversion problems with features
similar to those of this simple example.

While minimization of $J_{\rm ESI}$ might be tackled directly - by
alternately minimizations between $m$ and $f$, or by computing updates
for $m$ and $f$ simultaneously - such joint mimization performs
poorly, as \cite{YinHuang:16} has shown. The reason for this poor
performance is that $J_{\rm ESI}$ has dramatically different
sensitivity to $m$ versus $f$, especially for high frequency $f$, as
the reader will see below. Instead, a nested approach, in which $f$ is
eliminated in an inner optimization,
generally gives far better numerical performance.  This Variable
Projection Method (VPM) \cite[]{GolubPereyra:03} takes advantage of
$J_{\rm ESI}$ being quadratic in
$f$ to solve for $f$ given $m$, thus producing a reduced objective of
$m$ alone:
\begin{equation}
\label{eqn:red}
J_{\rm VPM}[m] = \min_f J_{\rm ESI}[m,f] = J_{\rm ESI}[m,f[m]],
\end{equation}
where $f[m]$ is the minimizer of $J_{\rm ESI}$ over $f$ for given $m$.
For the problem considered here, $J_{\rm VPM}$ is explicitly computable. First observe that apart from amplitude, $S[m]$ is unitary:
\begin{equation}
\label{eqn:tran}
S[m]^T g (t) = \frac{1}{4\pi r}g\left(t+mr\right)
\end{equation}
so
\begin{equation}
\label{eqn:unit}
S[m]^TS[m] f (t) = \frac{1}{(4\pi r)^2} f(t).
\end{equation}
Therefore the normal equation for the minimizer on the RHS of equation \ref{eqn:red} is
\begin{equation}
\label{eqn:norm1}
\left(\frac{1}{(4\pi r)^2} I + \alpha^2 A^TA\right)f[m] = S[m]^T\dl.
\end{equation}

At this point I have to come clean about the actual choice of
$A$. Recall that $A$ is called an ``annihilator'', as it vanishes when
applied to the target source field, at least in an idealized
limit. Thus the choice of $A$ depends on modeling assumptions, not
fundamental physics, as is characteristic of extended source
inversion. As $\lambda \rightarrow 0$, the target source wavelet $\fl$
focuses at time $t=0$, in the sense that it vanishes for
$|t| > \lambda$. Therefore I choose $A$ to penalize energy away from
$t=0$:
\begin{equation}
\label{eqn:ann}
Af(t) = tf(t).
\end{equation}
This
particular annihilator has been employed in earlier papers on extended
source inversion
\cite[]{Plessix:00a,LuoSava:11,Warner:14,HuangSymes:SEG15a,Warner:16,HuangSymes:GEO17}.

With this choice of $A$, the normal operator on the LHS of \ref{eqn:norm1} is simply multiplication by a positive function of time, and can be inverted by inspection. Using the identity \ref{eqn:tran} for the adjoint operator, obtain
\[
f[m](t) = \left(\frac{1}{(4\pi r)^2} + \alpha^2
  t^2\right)^{-1}\frac{1}{4\pi r}\dl\left(t+mr\right)
\]
\begin{equation}
\label{eqn:solnnon}
= \left(1+ (4\pi r)^2\alpha^2 t^2\right)^{-1}\fl\left(t+(m-m_*)r\right) 
\end{equation}
thanks to the definition of the data $\dl$ (equation \ref{eqn:defdata}).
After a little algebra,
\begin{equation}
\label{eqn:resid}
(S[m]f[m]-d)(t) =\frac{1}{4\pi r}[ (1+(4\pi r)^2 \alpha^2 (t-mr)^2)^{-1} - 1]\fl(t-m_*r)
\end{equation}
and
\begin{equation}
  \label{eqn:anniexp}
  Af[m] = t \left(1+ (4\pi r)^2\alpha^2
    t^2\right)^{-1}\fl\left(t+(m-m_*)r\right).
\end{equation}
So
\[
  J_{\rm VPM}[m] = \frac{1}{2}\int\,dt\, \left( \frac{1}{(4\pi r)^2}[
    (1+(4\pi r)^2 \alpha^2 (t+(m_*-m)r)^2)^{-1} - 1]^2 +
  \right.
  \]
  \[
   \left. \alpha^2(t+(m_*-m)r)^2 \left(1+ (4\pi r)^2\alpha^2
      (t+(m_*-m)r)^2\right)^{-2}\right)|\fl(t)|^2
\]
\begin{equation}
\label{eqn:vpmexpl}
=\frac{1}{2(4\pi r)^2}\int\,dt\, [1-(1+(4\pi r)^2 \alpha^2
(t+(m_*-m)r)^2)^{-1}]|\fl(t)|^2
\end{equation}

The gradient of $J_{\rm VPM}$ can be extracted by elementary means
from the identity \ref{eqn:vpmexpl}, but instead we will derive it in
using some important features that this inverse problem shares with
other inverse wave problems with more complex physics. To begin with, the gradient of a VPM
objective of the form \ref{eqn:red} is given by the formula 
\begin{equation}
\label{eqn:grad}
\nabla J_{\rm VPM}[m] = (DS[m]f[m])^*(S[m]f[m]-d).
\end{equation}
This easily derived result is in some sense the main content of \cite[]{GolubPereyra:03}.  In this formula, $DS[m]f$ is the derivative
of the modeling operator $S[m]f$
with respect to $m$, that is, with $f$ held fixed. The adjoint $(DS[m]f)^*$ is the adjoint of the map from model space to data space:
\[
(DS[m]f): \delta m \mapsto (DS[m]\delta m)f
\]
and is NOT the same as the adjoint denoted by $S[m]^T$, which is the source-space-to-data-space adjoint.
That is, this adjoint make this relation true, for all $m, \delta m,f,$ and $d$:
\begin{equation}
\label{eqn:vadj}
\delta m \cdot (DS[m]f)^*d = \int \,dt\,[(DS[m]\delta m)f] (t) d(t) 
\end{equation}
On the left side is the dot product in model space - since model space
is just 1D in this example, that's just the numerical product. On the
right is the dot product in data space, in the idealized continuum
limit.

Note that the VPM gradient formula is remarkable in two ways. First,
it is exactly the same as the FWI gradient, that is, the gradient of
$J_{\rm FWI}$ if you happen to insert the solution $f[m]$ of the
normal equation for $f$. Second, it does not mention the annihilator
$A$ at all: its impact is locked up in $f[m]$.

The key to unlocking the meaning of the VPM gradient formula for this
and similar problems is a remarkable feature of the derivative
operator $DS[m]$: from the definition \ref{eqn:mod},
\begin{equation}
\label{eqn:deriv}
(DS[m]\delta m)f (t) = -\frac{\delta m}{4\pi}\frac{df}{dt}(t-mr) = S[m](Q[m]\delta m)f (t),
\end{equation}
where 
\begin{equation}
\label{eqn:defq}
(Q[m]\delta m)f = -r\delta m \frac{df}{dt}.
\end{equation}
That is, $Q[m]\delta m$ is a skew-adjoint operator depending linearly
on $\delta m$ - more on this below.

A calculation, detailed in Appendix A, yields an expression for the
gradient \ref{eqn:grad} in terms of $Q$:
\begin{equation}
\label{eqn:gradcomm}
\delta m \cdot \nabla J_{\rm VPM}[m] = \frac{1}{2}\alpha^2\int \,dt\,f[m](t)[(Q[m]\delta m),A^TA]f[m](t)
\end{equation}
Here, the symbol $[L,M]$ denotes the {\em commutator} of the operators
$L$ and $M$: $[L,M]=LM-ML$.

Note that the annihilator $A$ is explicitly present in
\ref{eqn:gradcomm}. The structure displayed in this expression is
common to many other extended inversion methods. The extreme
simplicity of the factorization \ref{eqn:deriv}, \ref{eqn:defq} and
the gradient expression \ref{eqn:gradcomm} is
modified for more complex inversion problems by asymptotically
negligible corrections \cite[]{Symes:IPTA14,tenKroode:IPTA14,Symes:Madrid}.

%%%%%%%%%%%

Remember that $A^TA$ amounts to multiplying by $t^2$, and $Q$ is the scaled time derivative (equation \ref{eqn:defq}), so
\begin{equation}
\label{eqn:comm3}
[(Q[m]\delta m),A^TA]=-2r\delta mt
\end{equation}
Insert this identity into equation \ref{eqn:gradcomm} to obtain
\begin{equation}
\label{eqn:comm4}
\delta m \cdot \nabla J_{\rm VPM}[m] =  -r\delta m \alpha^2\int \,dt\,tf[m]^2(t)
\end{equation}
Combine this identity with the formula \ref{eqn:solnnon} for the
solution of the inner problem, and divide out the common factor
$\delta m $, to obtain
\begin{equation}
\label{eqn:gradfinal}
\nabla J_{\rm VPM}[m] = -r\alpha^2\int \,dt\,t\left(\frac{1}{(4\pi r)^2} + \alpha^2 t^2\right)^{-1}\frac{1}{(4\pi r)^2}\fl\left(t+(m-m_*)r\right)^2
\end{equation}
Recall that $f_1(s)$ vanishes if $|s|\ge 1$, so $\fl(t+(m-m_*)r)$
vanishes if $|t+(m-m_*)r| > \lambda$. Therefore the integral on the
RHS of equation \ref{eqn:gradfinal} can be re-written
\[
  = -r\alpha^2\int_{-(m-m_*)r-\lambda}^{-(m-m_*)r+\lambda}
  \,dt\, \frac{t}{1+(4\pi r)^2\alpha^2 t^2}\fl\left(t+(m-m_*)r\right)^2
\]
If $m > m_*+\lambda/r$, then the entire interval of integration is a proper
subset of the negative half-axis. Consequently the first factor in the
integrand satisfies
\[
 \frac{t}{1+(4\pi r)^2\alpha^2 t^2}
  \le -(m-m_*)r+\lambda < 0
\]
over the interval of integration. Consequently, the integral is
negative (since $\fl$ has a positive mean square) and so the gradient
is positive. Similar reasoning applies to the case $m <
m_*-\lambda/r$.

To summarize,
\begin{itemize}
\item if $m > m_*+\lambda/r$, then $\nabla J_{\rm VPM}[m] > 0$, and
\item if $m < m_*-\lambda/r$, then $\nabla J_{\rm VPM}[m] < 0$.
\end{itemize}
That is, {\em $J_{\rm VPM}$ has no local minima
  further than  $O(\lambda)$ from the global minimum:} the gradient
has the correct sign and slowness updates computed from it will be
constructive, unless the slowness estimate is already ``within a
wavelength'' of being correct.

On the other hand, careful examination of equation \ref{eqn:gradfinal}
shows that $J_{\rm VPM}$ is not convex: there is an inflection point
$O(1/\alpha)$ from the global minimizer $m_*$, and in fact as $\alpha
\rightarrow \infty$ for fixed $\lambda$, the $J_{\rm VPM}$
approximates $J_{\rm FWI}$. 

\section{Discussion}
There remain several important points to be made. 

First, the formal computations centering on the operator $Q$
(equations \ref{eqn:deriv} through \ref{eqn:gradcomm}) depend only on
the relation \ref{eqn:deriv} and the skew-symmetry of $Q$, which hold
with minor modifications for other more complex waveform inversion
problems. These other more complex problems do not submit to such a
simple treatment as is shown in the equations following
\ref{eqn:gradcomm}, but for example it is possible in some cases to
use a relation analogous to \ref{eqn:deriv} to extract a relation between extended waveform inversion and traveltime tomography, via analysis of the Hessian at a zero-residual global minimizer. See for instance \cite{tenKroode:IPTA14,Symes:IPTA14,Symes:Madrid}. Up to that point the reasoning is quite general, and central to the understanding gained so far of extended inversion methods. 

Another issue mentioned in the last section is that the penalty weight
$\alpha$ must be chosen somehow, and its choice influences heavily the
convergence rate of iterative optimization algorithms applied to $J_{
  \rm VPM}$ (or $J_{\rm ESI}$, though (as also mentioned above) it is a
poor candidate for local optimization). Two solutions to the weight
assignment problem has emerged in the last few years, both quite effective. One applies
directly to VPM problem, via use of the Discrepancy Principle
\cite[]{Morozov:84}. This algorithm controls the size of the data
misfit term in $J_{\rm ESI}$ to lie in an interval chosen to contain
the assumed level of data noise, by adjusting $\alpha$ sporadically as
the optimization proceeds. While this approach requires an estimate of
data misfit to be achieved at the optimal model, it is very effective
in maintaining convergence \cite[]{FuSymes2017discrepancy}. Another
approach modifies the VPM problem by means of the Augmented Lagrangian
method \cite[]{NocedalWright}. This reformulation appears to suppress
much of the sensitivity to $\alpha$ of VPM optimization, and has been
successfully used in extended inversion \cite[]{Aghamiry:19}.

Perhaps the most important general message implicit in the example
presented here is that the choice of the annihilator $A$ determines
whether the extended inversion algorithm achieves global or semiglobal
convexity, as is accomplished in the present example. The annihilator
used here has a property whose importance can be guessed by
examining the gradient formula \ref{eqn:gradcomm}. The operator $Q$ is
a first order differential operator. The gradient is a quadratic form
whose Hessian is the communator $[Q,A^TA]$, and whose argument is
$f[m]$. In order that the VPM objective be continuous for any model
$m$ and finite energy data $d$, this form should admit any
finite-energy source as argument: in technical terms, it should be a
bounded (or continuous) operator on the Hilbert space of finite energy
traces. If one asks for a bit more, namely that VPM objective function
have derivatives of arbitrary order, then it is not too hard to see
that the iterated commutators $[Q,...[Q,A^TA]...]$ must all be bounded
operators. This is a very strong restriction on $A^TA$: this operator
must be {\em pseudodifferential}, that is, a combination of
differential operators and powers of the Laplace operator
\cite[]{Tay:81}. For the present operator, $A^TA$ is
multiplication by $t^2$, a very simple pseudodifferential
operator. For more discussion of this requirement in the context of
annihilators in extended inverseion, see \cite{geoprosp:2008}, where
you can also find references to the technical backstory.

This constraint actually rules out some popular approaches to FWI. To
begin with, basic least-squares FWI as formulated in the third section
above can be reformulated as a quadratic form whose Hessian turns out
not to be pseudodifferential. Therefore the FWI objective is not
smooth in data and model jointly, a fact that is linked to the
cycle-skipping behaviour demonstrated above. More surprising, perhaps,
the same turns out to be true for Wavefield Reconstruction Inversion (WRI),
an extended source inversion algorithm introduced by
\cite{LeeuwenHerrmannWRI:13}, and further developed by
\cite{LeeuwenHerrmann:16}, \cite{WangYingst:SEG16}, and
\cite{Aghamiry:19}, amongst others. This approach turns out to be closely
linked to basic FWI, and can be formulated as minimization of a similar quadratic form
which once again has a non-pseudodifferential Hessian. Not
coincidentally, it also exhibits cycle-skipping behaviour. Just as for
FWI, in simple cases such as the problem studied in this paper, WRI
can be shown explicitly to have local minima far from the global
minimum, and a region of attraction for the global minimum on the
order of a wavelength in diameter, just as does FWI \cite[]{wwsorcas:20-01}.

\section{Conclusion}
Desipite its simplicity, the single-trace transmission inversion
problem proves typical of many more complex waveform inversion
problems. The structure of the derivative is similar in many of these
problems, and for the particularly simple one explained here, can be
analysed on paper to the point of showing explicitly why a simple
extended source approach to waveform inversion works - that is,
generates an objective all of whose local minima are ``within a
wavelength'' of the global minimizer. Otherwise stated, this
particular extended source inversion is genuinely immune to
cycle-skipping. The simple structure of this problem showcases the
importance of the variable projection reduction (elimination of the
extended source) and a proper choice of annihilator in the formulation
of the basic objective. It also makes clear the central role played by
a factorization of the linearized modeling operator, a feature shared
with many more complex extended source methods applicable at field scale.

\append{Calculation of the VPM Gradient}
Using equations \ref{eqn:deriv}, \ref{eqn:defq} and \ref{eqn:vadj},
and \ref{eqn:grad},
\[
\delta m \cdot \nabla J_{\rm VPM}[m] =  \int \,dt\,[(DS[m]\delta m)f[m]] (t) (S[m]f[m]-d)(t)
\]
\[
= \int \,dt\, [S[m](Q[m]\delta m)f[m](t)] (S[m]f[m]-d)(t) = \int
\,dt\,(Q[m]\delta m)f[m](t) S[m]^T(S[m]  f[m]-d)(t)
\]
Now invoke the normal equation \ref{eqn:norm1}: and replace the last factor:
\begin{equation}
\label{eqn:comm1}
= -\alpha^2\int \,dt\, (Q[m]\delta m)f[m](t)A^TAf[m](t)
\end{equation}
Since $Q[m]\delta m$ is skew-symmetric, shift it onto the other factor in this $L^2$ inner product (why not):
\[
= \alpha^2\int \,dt\, f[m](t)(Q[m]\delta m)A^TAf[m](t) 
\]
\begin{equation}
\label{eqn:comm2}
= \alpha^2\int \,dt\, (A^TAf[m](t)(Q[m]\delta m)f[m](t) + f[m](t)[(Q[m]\delta m),A^TA]f[m](t))
\end{equation}
where I swapped $Q$ and $A^TA$ at the cost of introducing a term involving the commutator $[Q,A^TA] = QA^TA-A^TAQ$, and rearranged the first term using the symmetry of $A^TA$. Now notice that this first term is exactly the same as the RHS of equation \ref{eqn:comm1}, except for the minus sign - so subtracting the RHS of \ref{eqn:comm2}
from \ref{eqn:comm1} and rearranging, get
\[
-\alpha^2\int \,dt\, (Q[m]\delta m)f[m](t)A^TAf[m](t) = \frac{1}{2}\alpha^2\int \,dt\,f[m](t)[(Q[m]\delta m),A^TA]f[m](t)
\]
hence 
\begin{equation}
\label{eqn:gradcomm-app}
\delta m \cdot \nabla J_{\rm VPM}[m] = \frac{1}{2}\alpha^2\int \,dt\,f[m](t)[(Q[m]\delta m),A^TA]f[m](t)
\end{equation}
which is identical to equation \ref{eqn:gradcomm}.
\bibliographystyle{seg}
\bibliography{../../bib/masterref}

\end{document}